\documentstyle[floats,multicol,aps,epsfig]{revtex}

\begin{document}
\twocolumn[\hsize\textwidth\columnwidth\hsize\csname @twocolumnfalse\endcsname

\title{Unconventional Transition from Metallic to Insulating 
Resistivity in the Spin-ladder Compound (Sr,Ca)$_{14}$Cu$_{24}$O$_{41}$  } 
\author{F.F. Balakirev $^{1,2}$,  J.B. Betts$^2$, G.S. Boebinger$^{1,2}$, N. Motoyama$^3$, H. Eisaki$^3$, S. Uchida$^3$.}
\address{$^1$Bell Laboratories, Lucent Technologies
Murray Hill, New Jersey 07974, USA\\
$^2$National High Magnetic
Field Laboratory, Los Alamos National Laboratory, 
MS E536, Los Alamos, New Mexico 87545, USA\\
$^3$Department of Superconductivity, University of Tokyo, 
Bunkyo-ku, Tokyo 113, Japan}
\maketitle

\begin{abstract}
Spin-ladder compounds make interesting analogs of the high-temperature superconductors\cite{c1},
because they contain layers of nearly one-dimensional ``ladders'' consisting of a square
array of copper and oxygen atoms. Increasing the number of legs in the ladders provides a
step-wise approach toward the two-dimensional copper-oxygen plane, that structure believed 
to be a key to high temperature superconductivity. Short-range spin correlations in ladders
have been predicted to lead to formation of hole pairs favorable for superconductivity,
once enough holes are introduced onto the ladders by doping\cite{c2}. Indeed, superconductivity
has been discovered in the two-leg ladder compound (Sr,Ca)$_{14}$Cu$_{24}$O$_{41}$  under high
pressure\cite{c3}. Here we show that charge transport in the non-superconducting state of 
(Sr,Ca)$_{14}$Cu$_{24}$O$_{41}$  shares three distinct regimes in common with high-temperature 
superconductors, including an unexplained insulating behavior at low temperatures 
in which the resistivity increases as the logarithm of the temperature. These 
observations suggest that the logarithmic divergence arises from a new localization 
mechanism common to the ladder compounds and the high-temperature 
superconductors\cite{c4,c5}, which may arise from nearly one-dimensional charge transport 
in the presence of a spin gap. 
\end{abstract}
]

It is increasingly believed that a successful theory of high-temperature superconductivity 
(HTS) must also account for the anomalous properties exhibited by the so-called ``normal 
state'', the resistive state observed at temperatures above the superconducting transition 
temperature, T$_c$. There are numerous similarities in the normal-state properties of HTS 
and ladder cuprates, including an energy gap for spin excitations and a crossover from 
insulator to metal upon hole doping. These suggest that the ladder compounds provide a 
valuable experimental laboratory for probing the unusual properties of the resistive normal 
state of the high-temperature superconductors.

\begin{figure}[tb]
\begin{picture}(1,1)(0,0)
\put (10,320){\LARGE \bf \sf a}
\put (0,195){\LARGE \bf \sf b} 
\put (120,190){\LARGE \bf \sf c}
\end{picture}
\epsfig{file=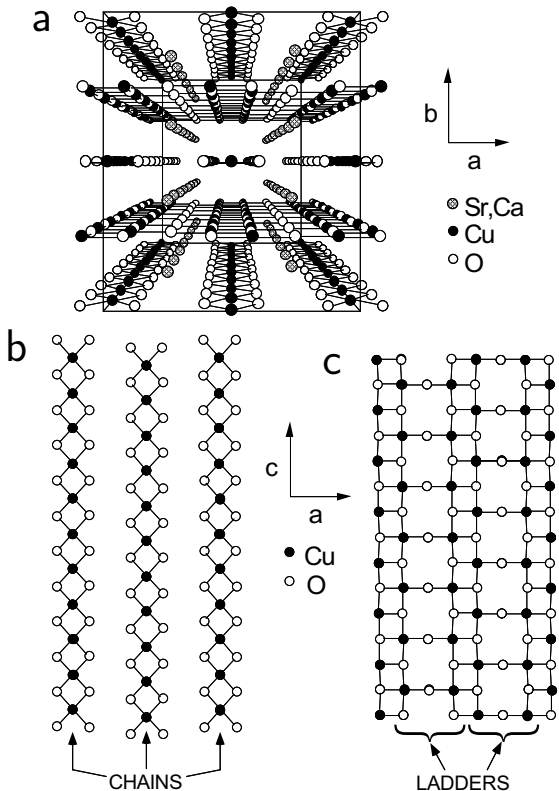,width=0.47\textwidth}
\caption{{\bf \sf a}, The crystal structure for (Sr,Ca)$_{14}$Cu$_{24}$O$_{41}$, including {\bf \sf b}, 
the plane containing CuO$_2$ chains and {\bf \sf c}, the plane containing Cu$_2$O$_3$ ladders.}
\label{fig_struct}
\end{figure}

In this letter we discuss resistivity measurements on five single-crystal samples of 
Sr$_{2}$Ca$_{12}$Cu$_{24}$O$_{41}$, which were grown by the traveling-solvent-floating-zone method\cite{c6}. As 
shown in Fig. \ref{fig_struct}, the crystal structure of this compound is composed of layers of CuO$_2$ 
chains and Cu$_2$O$_3$ two-leg ladders, interleaved with Sr$_{1-x}$Ca$_x$ buffer layers. The undoped 
parent compound, Sr$_{14}$Cu$_{24}$O$_{41}$, exhibits semiconducting behavior with an activation 
energy gap of 0.18eV\cite{c7}. The formal valency of Cu in Sr$_{14}$Cu$_{24}$O$_{41}$ is +2.25,
corresponding 
to a partially-filled valence band which ordinarily would result in metallic conductivity. 
However, in Sr$_{14}$Cu$_{24}$O$_{41}$, the positively-charged carriers (the ``holes'') are
located on the 
CuO$_2$ chains, which do not conduct because the transfer integral along the 90$^0$ oxygen 
bond is small (Fig. \ref{fig_struct}(b)). Upon partial substitution of Ca for the isovalent Sr (or upon the 
application of pressure), the holes are redistributed from the chains to the ladders, which 
are more conductive due to the 180$^0$ oxygen bonds\cite{c6}. This ``self-doping'' leads to a 
decrease of resistivity and, eventually, a crossover from insulating to metallic behavior as 
the carrier concentration is increased.  

\begin{figure}[tb]
\epsfig{file=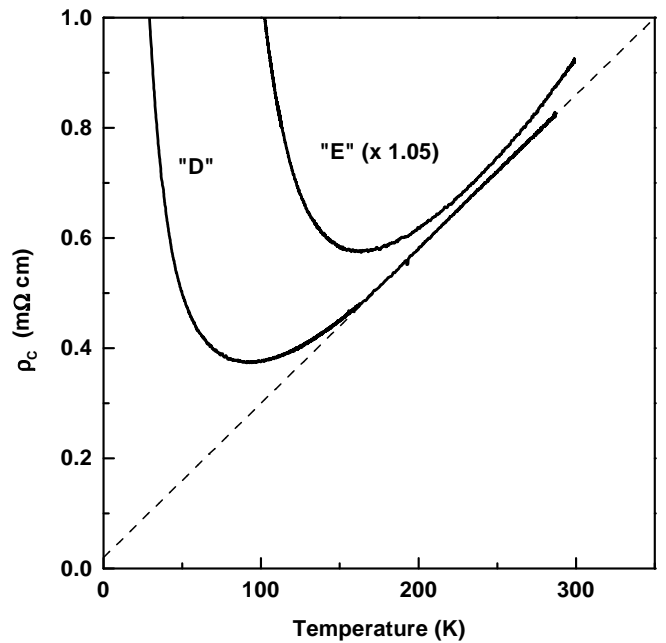,width=0.48\textwidth}
\caption{Resistivity along the ladder, $\rho_c$, for two Sr$_{2}$Ca$_{12}$Cu$_{24}$O$_{41}$   samples. Dashed line is the 
best linear fit for sample ``D'' at high temperatures.  The data for sample ``E'' is shifted for 
clarity.}
\label{fig_linear}
\end{figure}

Fig. \ref{fig_linear} presents c-axis resistivity, $\rho_c$, for two of the samples, denoted ``D''  and ``E'', 
measured using the standard four-probe method with the current applied along the ladder. 
In both of these samples, the temperature dependence of $\rho_c$ is metallic ($d\rho_c/dT >0$) at 
room temperature, although only in sample ``D''  does $\rho_c$  follow a linear dependence ($\rho_c \sim 
aT+b$) that extrapolates (dashed line) to a nearly-zero residual resistivity in the $T \rightarrow 0$ limit. 
A resistivity with a strictly linear temperature dependence is one of the striking 
unexplained trademarks of the normal state of the high-temperature superconductors\cite{c8}. The 
data of Fig. \ref{fig_linear} suggest that it is also a signature of transport in the two-leg ladder cuprate, 
once enough carriers are introduced onto the ladder.

\begin{figure}[tb]
\epsfig{file=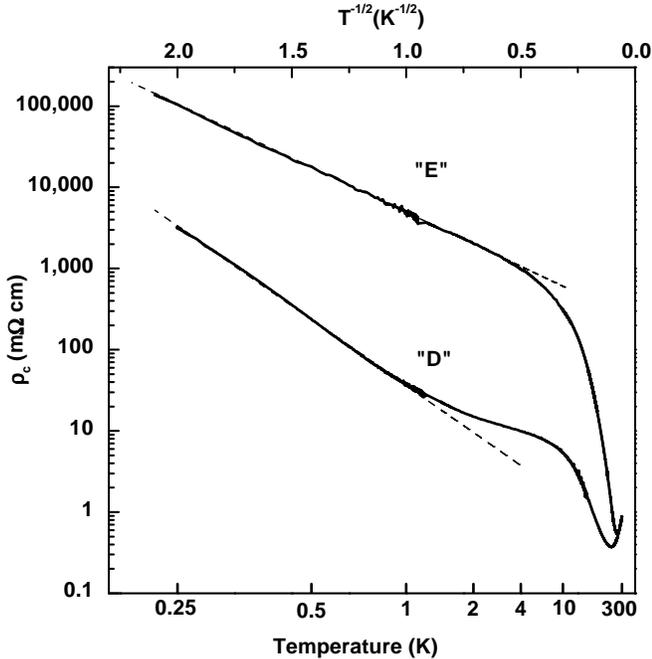,width=0.48\textwidth}
\caption{Variable range hopping (dashed lines) in the lowest temperature range for both 
samples.} 
\label{fig_vrh}
\end{figure}

At lower temperatures, both samples experience a crossover from metallic to insulating 
behavior ($d\rho_c/dT <0$). Although both samples have the same nominal composition, the 
insulating behavior begins at a higher temperature in sample ``E'' and the low-temperature 
resistivity is roughly two orders of magnitude larger in sample ``E'' than in sample ``D''. 
This difference can be due to different amounts of disorder and/or different levels of the 
carrier concentration between two samples. Nevertheless, in the extreme low-temperature 
limit, $\rho_c$ for both samples follows the temperature dependence expected for variable range 
hopping (VRH) of strongly localized carriers: $\rho_c = \rho_0\exp(T_0/T)^\beta$. The best fit to the data 
over the widest temperature range (Fig. \ref{fig_vrh}) corresponds to $\beta=1/2$, the same temperature 
dependence reported in highly resistive samples of the HTS cuprates\cite{c9,c10,c11}.  In this VRH 
regime, electrical current is dominated by variable-length hops of the charge carriers 
between localized states at the minima of the random disorder potential. The $\beta=1/2$ 
exponent can result from VRH in the presence of Coulomb repulsion between carriers, 
which suppresses the density of states at the Fermi energy in a low-carrier-density 
system\cite{c12}.

\begin{figure}[tb]
\epsfig{file=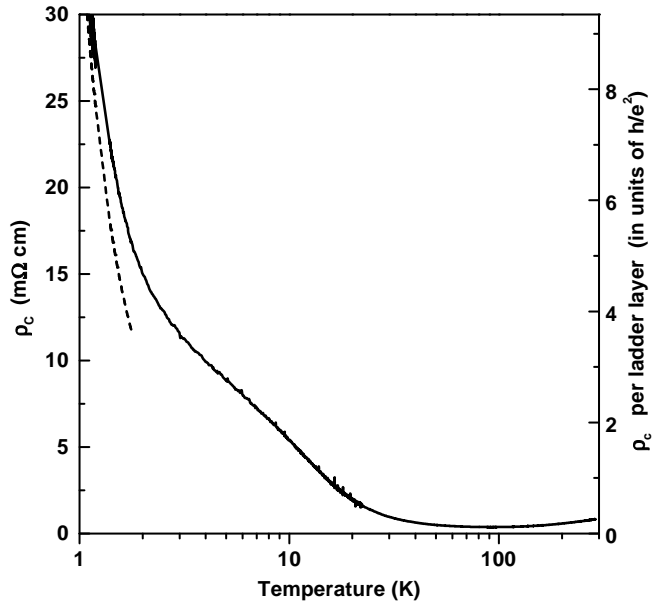,width=0.48\textwidth}
\caption{Intermediate temperature regime (3K $< T <$ 20K) in sample ``D'', which is best 
described by $\rho_c(T) \sim \log (1/T)$. Dashed line is the extrapolation of the VRH fit from Fig. \protect\ref{fig_vrh}. 
Right axis shows normalized  $\rho_c$ per layer of Cu$_2$O$_3$ ladders, in units of the quantum 
resistance, h/e$^2$.} 
\label{fig_loglin}
\end{figure}

The most surprising feature of the data occurs in an intermediate temperature regime 
between ~3K and ~20K in sample ``D'' (Fig. \ref{fig_loglin}). This regime is characterized by a 
temperature dependence best approximated by a logarithmic insulating behavior, $\rho_c \sim 
\log (1/T)$ \cite{c13}. In this log-$T$ regime, the magnetoresistance (MR) of sample ''D'' 
(measured at $T=4.2$K with the magnetic field applied along the b-axis) was found to be 
negative, while in the strong localization regime (measured at $T=1.2$K) the MR was found 
to be positive, as is typical for VRH. This gives additional experimental evidence that 
there is a third transport regime in sample ``D'', distinctly different from the linear-$T$ and 
strong localization regimes.

Of the three transport regimes, the lowest-temperature strong localization regime is the 
most robust, since VRH is observed at temperatures below 2K in every sample. This is the 
same temperature range in which long range spin order has been reported in 
(Sr,Ca)$_{14}$Cu$_{24}$O$_{41}$, in which spins on neighboring Cu-ladder sites are anti-aligned\cite{c14}.
It is consistent with our observations to suppose that the onset of long range anti-ferromagnetic
order gives rise to the onset of strong localization of charge carriers in the 
ladders. 

The other two transport regimes observed in sample ``D'' are not nearly so robust, perhaps 
signaling a greater sensitivity to disorder for the underlying transport mechanisms. After 
all, quasi-1D transport would be particularly sensitive to disorder: a mobile charge on a 
ladder may well have difficulty getting past a defect or break in that ladder. The cleanest 
linear-$T$ dependence is observed in sample ``D'', while samples with higher room-
temperature resistivity tend to show quasi-linear dependence of $\rho_c$ for which the 
extrapolated zero-temperature resistivity is non-zero. In samples with sufficiently high 
room-temperature resistivity ($>2$m$\Omega$ cm), insulating behavior ($d\rho_c/dT <0$) is observed 
even at room temperature, probably due to larger amounts of disorder in these samples. 
Although evidence of the log-$T$ regime has been found in two samples, we note that the 
unambiguous observation of the log-$T$ regime occurs in the sample which exhibits the 
cleanest linear-$T$ regime ($\rho_c  \sim aT$) of all samples studied.

Like the linear-$T$ transport regime, a log-$T$ transport regime also exists in the normal state 
of the HTS cuprates. There is growing experimental evidence that the anomalous 
normal-state properties of the HTS are due to the strong electron interactions near an 
insulator-to-metal crossover. The crossover is ordinarily obscured in HTS by the appearance of the 
superconducting phase; however, by suppressing superconductivity with an intense, pulsed 
magnetic field, the insulator-to-metal crossover in La$_{2-x}$Sr$_x$CuO$_4$ has been found to occur 
near optimum doping\cite{c15}, that carrier density which yields the maximum T$_c$. Underdoped 
samples, those with fewer carriers than optimal doping, exhibit a transport regime 
characterized by a log-$T$ divergence of the normal state resistivity, once superconductivity 
is lifted by the magnetic field\cite{c4,c5}. This divergence is inconsistent with known models for 
log-$T$ insulating behavior: it is apparently not arising from weak localization due to 
coherent backscattering\cite{c4,c5} or disorder-enhanced electron interactions, neither is it likely 
due to spin-flip (Kondo) scattering\cite{c4,c5}.

While there are differences between the CuO$_2$ plane of HTS cuprates and the plane 
containing 2-leg ladders in Sr$_{2}$Ca$_{12}$Cu$_{24}$O$_{41}$, we note that the log-$T$ insulating behavior 
occurs in both systems at the same magnitude of normalized resistivity, when the 
resistivity per layer is near the quantum resistance, h/e$^2 \sim 25.8$k$\Omega$. In a conventional 
two-dimensional system the quantum resistance corresponds to that resistance at which the 
mean free path is comparable to the deBroglie wavelength at the Fermi energy, which is 
where transport typically crosses from metallic (diffusive) to insulating (localized) 
behavior.
It is tempting to suggest that similar physical mechanisms can govern normal-state 
transport properties in the two-leg ladder compound and the HTS cuprates, even though, 
prima facia, the former contains quasi-one-dimensional transport along Cu$_2$O$_3$ ladders, 
while the HTS cuprates contain quasi-two-dimensional transport in the CuO$_2$ plane. 
Nonetheless, the phenomenology of three different transport regimes is similar in the two 
systems, including the two regimes (linear-$T$ and log-$T$) for which the underlying physical 
mechanisms remain unknown. In light of the data presented here, when coupled with 
published evidence of charged-stripe formation in HTS\cite{c17}, one could speculate that the 
effective dimensionality of the HTS CuO$_2$ plane is reduced with regard to charge transport 
and that the charge-transport mechanisms are similar in the two systems.


\begin{references}
\bibitem{c13}
  We note that the data in Fig. \ref{fig_loglin} has a slight curvature in the semi-log plot suggesting a 
dependence somewhat weaker than purely logarithmic. This curvature is reproducible and larger 
than any known measurement errors.
\end{references}
\end{document}